\documentclass[aps,pra,preprint,groupedaddress]{revtex4-2}

\usepackage{graphicx,caption,subcaption,amsmath}
\renewcommand{\varphi}{\wp} 

\begin{document}


\title{On hyperbolic and rational solutions of the cubically nonlinear Schr\"odinger equation}


\author{Hans Werner Sch\"urmann}
\email[]{hwschuer@uos.de}
\affiliation{Department of Mathematics, Computer Science, and Physics\\ University of Osnabr\"uck, Germany}

\author{Valery Serov}
\email[]{vserov@cc.oulu.fi,valserov@gmail.com}
\affiliation{Research Unit  of Mathematical Sciences\\ University of Oulu, Finland}


\begin{abstract}
In a previous article we have proved non-existence of certain "solutions" of the cubically nonlinear Schr\"odinger 
equation in the general case, 
and presented solutions in the non-generic case. -- In the present article we describe  a further family of solutions enlarging 
the set of non-generic solutions.  
\end{abstract}


\maketitle


\section{Introduction}

Forty years ago, in a seminal article \cite{AEK}, Akhmediev, Eleonskii, and Kulagin proposed  the solution-ansatz
$$
 \Psi(x,t)=(Q(x,t)+i\delta(t))e^{i\phi(t)},\quad Q, \delta,\phi\in \bf{R}
$$
for the nonlinear Schr\"odinger equation (NLSE)
$$
i\Psi(x,t)+\Psi_{xx}(x,t)+2|\Psi(x,t)|^2\Psi(x,t)=0.
$$
In a previous article \cite{Se} (cited as I in the following) we criticized \cite{AEK} by pointing out that Eq.(5) in \cite{AEK}, 
equivalent with Eq.(14a) in I, is violated in the general case (arbitrary integration constants (parameters) 
$a, c_1, c_2, c_3, h_0, f_0,$ related to the integration constants $W, H, D$ in \cite{AEK}). We did not notice in I, 
that for particular parameters Eq.(14a) in I can be satisfied, debunking the set of non-generic solutions in I as too restricted. 

After the statement of the problem in Section II, we present sufficient conditions for certain solutions of the NLSE in Section III. 
We apply the conditions to particular cases in Section IV.  The concluding Section V consists of a summary and remarks.

\section{Statement of the problem}

As outlined in I, ansatz
\begin{equation}
 \Psi(t,z)=(f(t,z)+id(z)e^{i\phi(z)},\quad f,d,\phi \in \bf{R}
\end{equation}
substituted in the cubically nonlinear Schr\"odinger equation
\begin{equation}
i \Psi_z (t,z)+\Psi_{tt}(t,z)+a\Psi(t, z)|\Psi(t, z)|^2=0,\quad a\in \bf{R}
\end{equation}
leads to two coupled nonlinear differential equations of Weierstrass type
\begin{equation}
(h_z(z))^2=\alpha_1h^4(z)+4\beta_1h^3(z)+6\gamma_1h^2(z)+4\delta_1h(z)+\epsilon_1=:R_1(h),
\end{equation}
where $h(z)=d^2(z)$ and
$$
\alpha_1=-16a^2,\quad \beta_1=4ac_1,\quad \gamma_1=-\frac{1}{3}(2c_1^2+8ac_2),\quad \delta_1=c_3,\quad \epsilon_1=0,
$$
and
\begin{equation}
(f_t(t,z))^2=\alpha_2f^4(t,z)+4\beta_2f^3(t,z)+6\gamma_2f^2(t,z)+4\delta_2f(t,z)+\epsilon_2=:R_2(f,z),
\end{equation}
where 
$$
\alpha_2=-\frac{a}{2},\quad \beta_2=0,\quad \gamma_2=\frac{1}{6}(c_1-3ah(z)),
$$
$$
\delta_2=-\frac{1}{2}\sqrt{c_3-(c_1^2+4ac_2)h(z)+4ac_1h^2(z)-4a^2h^3(z)},\quad
\epsilon_2=2c_2-c_1h(z)+\frac{3}{2}ah^2(z).
$$
The solutions are given by Eq.(11) in I. The coefficients in the quartic polynomial $R_1(h)$ depend on the parameter $a$ and 
the integration constants $c_1,c_2,c_3$ (see (9) in I); Eqs.(3) and (4) are coupled by the coefficients of the quartic polynomial 
$R_2(f,z)$ via the solution $h(z)$ of (3) and $c_1,c_2,c_3$ (see Eq.(10) in I). The problem is: Can parameters $a, c_1, c_2, c_3$, 
initial value $h_0$,  and initial function $f_0(z)$ be chosen such that system (14) in I
\begin{equation}
T:=f_z(t,z;f_0(z))-sign(z)\sqrt{h(z)}(c_1-3ah(z)-af^2(t,z;f_0(z))),
\end{equation}
\begin{equation}
(f_t(t,z;f_0(z)))^2=R_2(f,z)
\end{equation}
is satisfied. We note that Eq.(6) is satisfied by solution (13) in I for any real $h_0, f_0(z) a, c_1,c_2,c_3$ due to the equivalence 
of $R(y)$ and $y(x)$ (see Eq.(11) in I).

\section{sufficient conditions for particular solutions of the NLSE}

We consider particular real, non-negative, and bounded solutions $h(z)$ of Eq.(3)
associated to phase diagrams $\{h^2_z, h\}$ represented in Fig.1. The family of solutions
is algebraically defined by
\begin{equation}
c_1^2=16ac_2,\quad c_3=2c_1c_2>0.
\end{equation}
Choosing the initial condition
\begin{equation}
h_0=h(0)=0,
\end{equation}
the solutions $h(z)$ are given by 
\begin{equation}
h(z)=\frac{24c_1c_2}{5c_1^2+\sqrt{12g_{2h}}
\left(1+3\sinh^{-2}(\sqrt{3\sqrt{\frac{g_{2h}}{12}}}z)\right)}.
\end{equation}
The invariants and discriminant of the quartic $R_1(h)$ are (due to (7))

\begin{equation}
g_{2h}=\frac{c_1^4}{12},\quad g_{3h}=-\frac{c_1^6}{256}\quad
\Delta_h=0.
\end{equation}
In the general case, solutions $h(z)$ of Eq.(3) are expressed in terms
of Weierstrass' function $\wp(z; g_{2h},g_{3h})$ ($\{g_{2h},g_{3h},\Delta_h\}$ classify both $R_1(h)$ and
$\wp(z; g_{2h},g_{3h})$ \cite{Note}). With condition (7), the general solution $h(z)$ of Eq.(3) degenerates
to the hyperbolic Eq.(9), which, with Eq.(8a) in I, leads to the phase function
\begin{equation}
\phi(z)=\frac{c_1z}{2}+\arctan(\tanh\left(\frac{c_1z}{2}\right)).
\end{equation}
Hence, function $f(t, z)$ in ansatz (1) remains to be determined. It is given by Eq.(13) in I
$$
f(t, z; f_0(z))=
$$
\begin{equation}
\frac{-2\gamma_2\delta_2 - (5\gamma_2^2-\alpha_2\epsilon_2)f_0(z) + 2\alpha_2\delta_2 f_0^2(z) + 4\wp(t)(\delta_2+2\gamma_2f_0(z)+\wp(t)f_0(z)) + 
2\wp_t(t)\sqrt{R_2(f_0(z), z)}}{(2\wp(t)-\gamma_2-\alpha_2f_0^2(z))^2-\alpha_2R_2(f_0(z), z)},
\end{equation}
where 
the invariants and discriminant of $f(t,z)$ (and $R_2(f,z)$ \cite{Note}) are (subject to (7))
\begin{equation}
g_{2t}=\frac{c_1^2}{48},\quad g_{3t}=\frac{c_1^3}{1728},\quad \Delta_f=0.
\end{equation}
Since $\Delta_f=0$, Weierstrass' function $\wp$ in Eq.(12) reduces to hyperbolic expressions
$$
\wp(t;g_{2t},g_{3t})=\sqrt{\frac{g_{2t}}{12}}
\left(1+3\sinh^{-2}(\sqrt{3\sqrt{\frac{g_{2t}}{12}}}z)\right), \quad c_1<0,
$$
$$
\wp(t;g_{2t},g_{3t})=-\sqrt{\frac{g_{2t}}{12}}
\left(1-3\sinh^{-2}(\sqrt{3\sqrt{\frac{g_{2t}}{12}}}z)\right), \quad c_1>0.
$$
Solutions $h(z)>0$ are real and bounded. To ensure the same for $f(t,z)$ we impose

\begin{equation}
R_2(f_0(z), z)=0
\end{equation}
and require the denominator of Eq.(12) to be non-vanishing. This leads to the condition
\begin{equation}
\alpha_2R_2(f_0(z),z)-\left(\alpha_2f_0^2(z)+\gamma_2(z)-2\wp(t,g_{2t},g_{3t})\right)^2 \ne 0.
\end{equation}
Thus, due to (14), the denominator does not vanish if, for all $z$,
\begin{equation}
\alpha_2f_0^2(z)+\gamma_2(z)-2\wp(t,g_{2t},g_{3t})\ne 0.
\end{equation}
Using the lower bound of the Weierstrass function, $\wp(t,g_{2t},g_{3t})\ge \sqrt{\frac{g_{2t}}{12}}$, a sufficient condition is
\begin{equation}
K:=\alpha_2f_0^2(z)+\gamma_2(z) - \sqrt{\frac{g_{2t}}{12}}<0,
\end{equation} 
which restricts not only possible combinations of the parameters but also the admissible roots $f_0(z)$ of Eq.(14).

The singularities of $\wp(t,g_{2t},g_{3t})$ in the numerator and the denominator cancel, and do not effect the solution $f(t,z)$.

For evaluation of $f(t,z)$ according to Eq.(13) we require existence of a simple (double roots imply $f_z(t,z)=0$ and can 
be disregarded here, since this case has been investigated in I (see Eqs.(27)-(30) in I) root of Eq.(14). Necessary for 
this requirement is $R_2(f_0(0),0)=0$, written as 
\begin{equation}
-\frac{a}{2}f_0^4(0)+c_1f_0^2(0)-2\sqrt{c_3}f_0(0)+2c_2=0.
\end{equation}
Due to (7), the discriminant of (18) vanishes. The cubic resolvent of (18) is given by 
\begin{equation}
Res = y^3 - \frac{4c_1^2}{a}y^2 +\left(\frac{4c_1^2}{a^2}+\frac{16c_2}{a}\right)y-\frac{16c_3}{a^2}.
\end{equation}

Under condition (7) the discriminant of (19) vanishes, and (19) has two real roots, one of which is double (a triple root does 
not exist, due to (7)). Hence, condition (19), necessary for existence of a real root $f_0(z)$ of $R_2(f_0(z),z)=0$, 
has four real roots $f_{01}(z),f_{02}(z),f_{03}(z),f_{04}(z)$, one of which is double (due to $\Delta_f=0$), two are simple. 

Subject to condition (7), the simple roots of $R_2(f_0(0),0)=0$ are ($c_1c_2>0$)
\begin{equation}
\text{if} \quad c_2>0,\quad f_{01}(0)=-2(\sqrt{2}-2)\sqrt{\frac{c_2}{c_1}},\quad f_{03}(0)=-2(\sqrt{2}+2)\sqrt{\frac{c_2}{c_1}},
\end{equation}
\begin{equation}
\text{if} \quad c_2<0,\quad f_{01}(0)=2(\sqrt{2}-2)\sqrt{\frac{c_2}{c_1}},\quad f_{03}(0)=-2\sqrt{2}\sqrt{\frac{c_2}{c_1}}.
\end{equation}
As mentioned above, condition (17) is suitable to select the appropriate $f_0(z)$ for evaluation of $f(t,z)$ according to Eq.(13) in I. 
Substituting of $f_{01}(0)$ and $f_{03}(0)$ into Eq.(17), yields
\begin{equation}
\text{if} \quad c_2>0,\quad K_{01}=\frac{1}{12}(6\sqrt{2}-7-c_1)c_1,\quad K_{03}=-\frac{1}{12}(6\sqrt{2}+7+c_1)c_1,
\end{equation}
\begin{equation}
\text{if} \quad c_2<0,\quad K_{01}=\frac{1}{12}(6\sqrt{2}-7-c_1)c_1,\quad K_{03}=-\frac{1}{12}(c_1+1)c_1,
\end{equation}
and thus the admissible criteria
\begin{equation}
K_{01}<0\quad \text{if}\quad c_1<0\quad \text{or}\quad c_1>6\sqrt{2}-7,
\end{equation}
\begin{equation}
K_{03}<0\quad \text{if} \quad c_1>0 \quad \text{or}\quad c_1<-1.
\end{equation}
According to (24) and (25), we must choose $f_0(z)$ as 
\begin{equation}
f_0(z)=2\sqrt{\frac{c_2}{c_1}}\left(\frac{\sqrt{1+\cosh(c_1z)}-2}{\sqrt{\cosh(c_1z)}}\right),\quad c_1<0,
\end{equation}
\begin{equation}
f_0(z)=-2\sqrt{\frac{c_2}{c_1}}\left(\frac{\sqrt{1+\cosh(c_1z)}+2}{\sqrt{\cosh(c_1z)}}\right),\quad c_1>0.
\end{equation}
With respect to the problem stated in Section II we summarize: If parameters $a, c_1, c_2, c_3$, initial value $h(0)$, and 
initial function $f_0(z)$ are selected according to 
$$
d(0)=0,\quad \bf{(C_1)}
$$
$$
c_1^2=16ac_2, c_3=2c_1c_2>0,\quad \bf{(C_2)}
$$ 
$$
f_0(z)=(26)\quad \text{if} \quad c_1<0, \quad\text{and} \quad f_0(z)=(27)\quad \text{if} \quad c_1>0,\quad \bf{(C_3)},
$$
the NLSE has solutions $\Psi(t,z)$ given by Eq.(1).

\section{applications of conditions $\bf{(C_1)},\bf{(C_2)},\bf{(C_3)}$} 

In order to test effectiveness of the foregoing conditions, we present examples in this Section.

1. \quad We choose in Eq.(2) $a=1,c_1=2,c_2=\frac{1}{4}, c_3=1$.  From Eq.(10) we obtain 
\begin{equation}
h(z)=\frac{\sinh^2(z)}{\cosh(2z)},
\end{equation}
so that $\bf{(C_1)}$ is satisfied. Obviously, $\bf{(C_2)}$ is satisfied also.
Equation (5) must be tested subject to $f_0(z)$ given by Eq.(27) (since $c_1>0$). 
Simplification of Eq.(27) yields

$$
f_{0}(z)=-\frac{\cosh(z)+\sqrt{2}}{\sqrt{\cosh(2z)}}.
$$
The corresponding solution $f(t,z)$ according to Eq.(13) can be simplified to
\begin{equation}
f(t,z)=\frac{\cos t\cosh z+\sqrt{2}\sinh^2z}{\sqrt{\cosh 2z}(\cos t-\sqrt{2}\cosh z)},
\end{equation}
which satisfies $f(0,z)=f_0(z)$.
Evaluating (analytically) Eq.(5) by using (28) and (29), the result is $T=0$, confirming consistency.

The phase function $\phi(z)$ is given by Eq.(11)
\begin{equation}
\phi(z)=z+\arctan(\tanh(z)).
\end{equation}
The resulting solution of the NLSE reads
\begin{equation}
\Psi(t,z)=\frac{(\cosh(z)+i\sinh(z))\left(\frac{\cos(t)\cosh(z)+\sqrt{2}\sinh^2(z)}{\sqrt{\cosh(2z)}(\cos(t)-\sqrt{2}\cosh(z))}-i\frac{\sinh(z)}{\sqrt{\cosh(2z)}}\right)}{\sqrt{\cosh(2z)}}e^{iz},
\end{equation}
which is simplified to
\begin{equation}
\Psi(t,z)=\frac{\cos(t)+i\sqrt{2}\sinh(z)}{\cos(t)-\sqrt{2}\cosh(z)}e^{iz},
\end{equation}
named the "Akhmediev-breather" in the literature \cite{AkhAT}.

The foregoing results are illustrated (Figs.1-3, 5) and confirmed (Fig.4) numerically.

\begin{figure}[h!]
\centering
\includegraphics[scale=1.4]{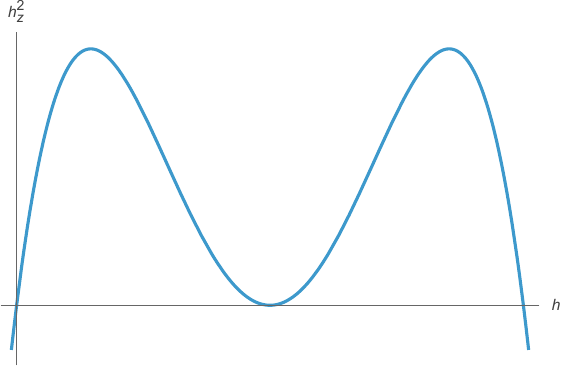}
\caption{Function $(h_z(z))^2(h)$ according to Eq.(3) with $\bf{(C_2)}$ satisfied}
\end{figure}

\begin{figure}[h!]
\centering
\includegraphics[scale=1.4]{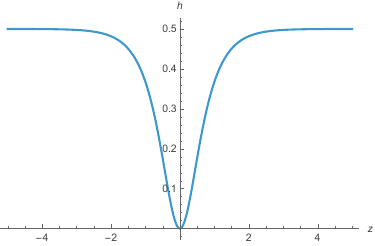}
\caption{$h(z)$ according to Eq.(28)}
\end{figure}

\begin{figure}[h!]
\centering
\includegraphics[scale=1.4]{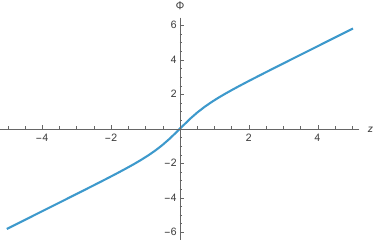}
\caption{$\phi(z)$ according to Eq.(30)}
\end{figure}

\begin{figure}[h!]
\centering
\includegraphics[scale=1.4]{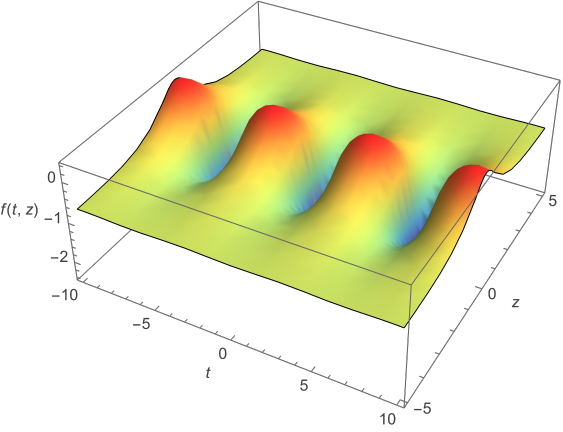}
\caption{$f(t,z)$ according to Eq.(29)}
\end{figure}

\begin{figure}[h!]
\centering
\includegraphics[scale=1.4,trim={2cm 17cm 7cm 3cm},clip]{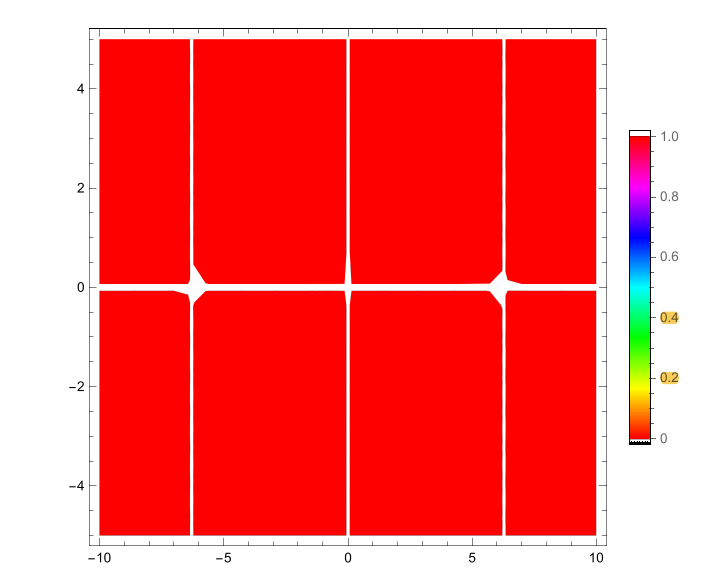}
\caption{Test of Eq.(5); $a=1, c_1=2, c_2=\frac{1}{4}, c_3=1, h(z)$ according to Eq.(10)}
\end{figure}

\begin{figure}[h!]
\centering
\includegraphics[scale=1.4]{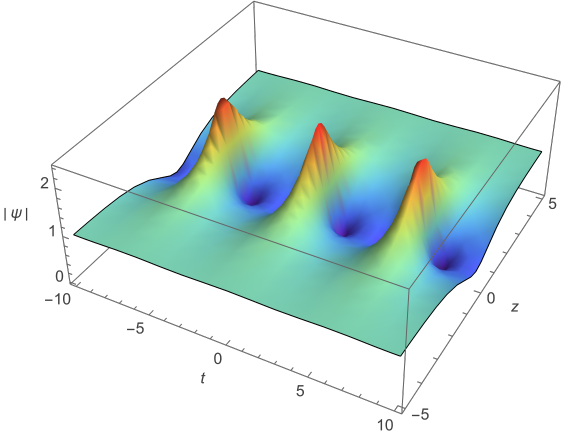}
\caption{$|f(t,z)+i\sqrt{h(z)}|$ according to Eq.(1)}
\end{figure}

\begin{figure}[h!]
\centering
\includegraphics[scale=1.4,trim={2cm 17cm 7cm 3cm},clip]{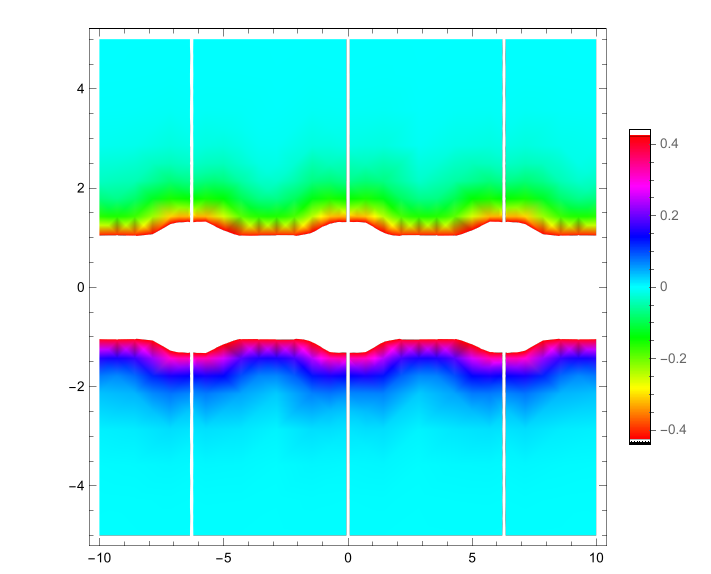}
\caption{Test of Eq.(5); parameters as in Fig.4, $h(z)$ according to Eq.(12) in I with $h_0=1$}
\end{figure}

2.\quad With the same parameters as in the foregoing example, if $h_0=1$, Eq.(13) in I yields 
\begin{equation}
h(z)=\frac{\cosh^2(z)}{\cosh(2z)}.
\end{equation}
Conditions $\bf{(C_1)},\bf{(C_2)}$ are satisfied, $\bf{(C_3)}$ not (Eq.(28) represents a dark-solitary solution, 
Eq.(33) a bright-solitary one). 
Using solution (33), numerical evaluation of Eq.(5) yields $T\ne 0$ (Fig.6), consistent with the general claim in I.

3.\quad Considering parameters $a=\frac{1}{8}, c_1=1, c_2=\frac{1}{2}, c_3=1$ and $h_0=0$, conditions 
$\bf{(C_1)},\bf{(C_2)},\bf{(C_3)}$ are satisfied, and we get
\begin{equation}
h(z)=\frac{4\sinh^2(z/2)}{\cosh(z)},
\end{equation}
\begin{equation}
f_0(z)=-\frac{\sqrt{2}(2+\sqrt{1+\cosh(z)})}{\cosh(z)},
\end{equation}
and a lengthy expression for $f(t,z)$. Numerical evaluation of Eq.(5)
yields $T=0$ (Fig.7), function $|\Psi(t,z)|$ is depicted in Fig.8.

4. \quad If we select $a=-1, c_1=-2, c_2=-\frac{1}{4}, c_3=1$ and $h_0=0$ (see example 1.), following the line as in the foregoing 
examples, we get $T=0$ in Eq.(5) (see Fig.9) and function $|\Psi(t,z)|$ depicted  in Fig.10.

5. \quad Finally, in view of possible applications in hydromechanics (ocean rogue-waves) and nonlinear optics 
(optical pulses), we present expressions for the maximal amplitude of the solutions of Section III.

The maximum of $f(t, z))$ is at $t=0,z=0$. Subject to $\bf{(C_1)},\bf{(C_2)}$ we obtain with $\bf{(C_3)}$
\begin{equation}
f_{\max}=2(\sqrt{2}-2)\sqrt{\frac{c_2}{c_1}},\quad c_1<0,
\end{equation}
\begin{equation}
f_{\max}=-2(\sqrt{2}+2)\sqrt{\frac{c_2}{c_1}},\quad c_1>0,
\end{equation}
For the Akhmediev - breather (example 1.) Eq.(37) yields $|f_{\max}|\approx 2.41$, Eq.(36) yields $|f_{\max}|\approx 0.41$. 
The corresponding values for example 3. are 4.82 and 0.82, respectively.

\begin{figure}[h!]
\centering
\includegraphics[scale=1.4,trim={2cm 17cm 7cm 3cm},clip]{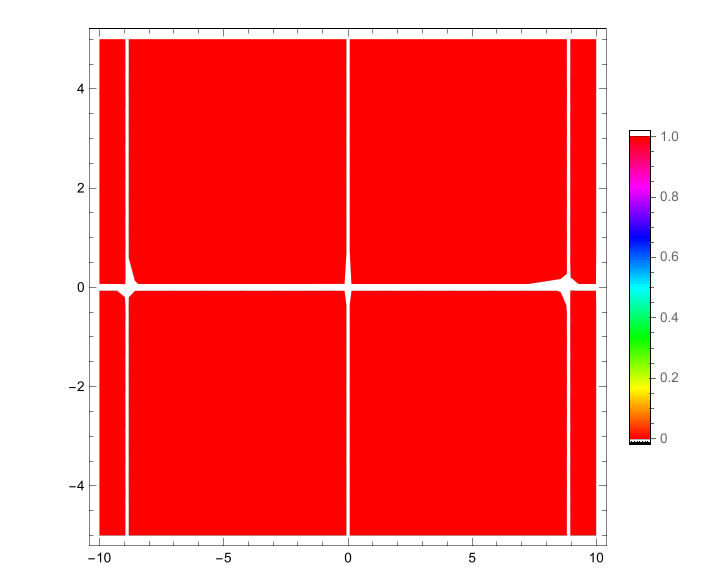}
\caption{Test of Eq.(5); $a=\frac{1}{8}, c_1=1, c_2=\frac{1}{2}, c_3=1, h(z)$ according to Eq.(10)}
\end{figure}

\begin{figure}[h!]
\centering
\includegraphics[scale=1.4]{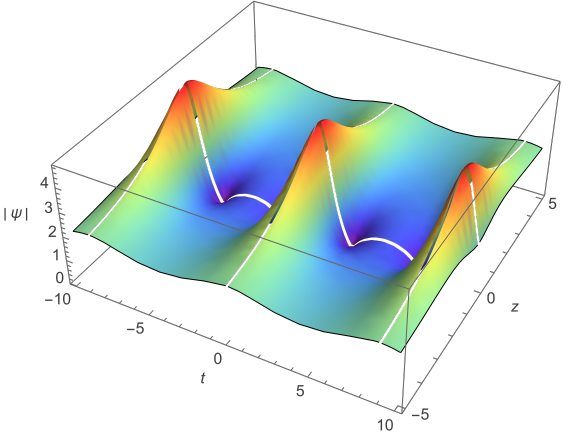}
\caption{$|\Psi(t, z)|; a=\frac{1}{8},c_1=-2, c_2=-\frac{1}{4}, c_3=1$}
\end{figure}

\begin{figure}[h!]
\centering
\includegraphics[scale=1.4,trim={2cm 17cm 7cm 3cm},clip]{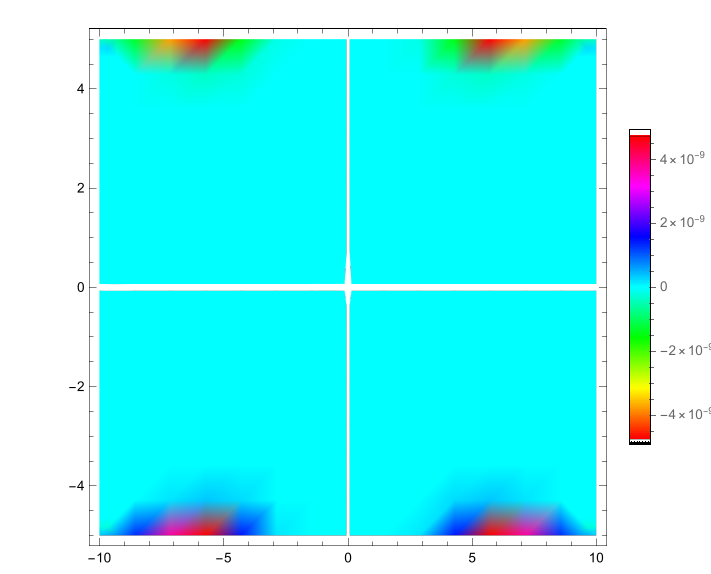}
\caption{Test of Eq.(5); $a=-1, c_1=-2, c_2=-\frac{1}{4}, c_3=1, h(z)$ according to Eq.(10)}
\end{figure}

\begin{figure}[h!]
\centering
\includegraphics[scale=1.4]{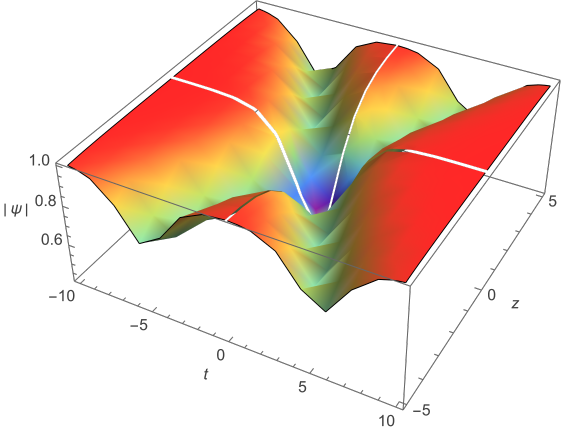}
\caption{$|\Psi(t, z)|; a=-1, c_1=-2, c_2=-\frac{1}{4}, c_3=1$}
\end{figure}


\begin{figure}[h!]
\centering
\includegraphics[scale=1.4]{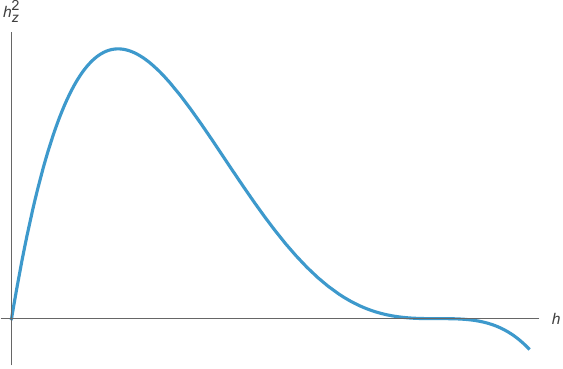}
\caption{Function $(h_z(z))^2(h)$ according to Eq.(3) with $\bf{(C^{*}_2)}$ satisfied,
representing rational solution $h(z)$}
\end{figure}

\section{Summary and remarks}

With respect to the problem posed in Section II, we have derived condition $\bf{(C_2)}$ together with the initial conditions 
$\bf{(C_1)}$ and $\bf{(C_3)}$ by considering a particular family of solutions $h(z)$ (Eq.(9)), represented by a phase diagram depicted in Fig.1, 
and by requiring $f(t, z)$ being real and bounded.

In Section IV, we have presented two examples showing that $\bf{(C_1)}$, $\bf{(C_2)}$, $\bf{(C_3)}$ are sufficient for 
$T=0$ in Eq.(5).

To conclude, we remark that conditions $\{\bf{(C_1)}, \bf{(C_2)},\bf{(C_3)}\}$ are defining a subspace $P_C$ in parameter space 
$P=\{a, c_1, c_2, c_3, h(0), f_0\}$.
It seems that (all) parameters in $P_C$ satisfy $T=0$ in Eq.(5). In favor of this conjecture we note that the conditions and $T$ 
are "analytic" with respect to the parameters (admittedly, a proof of the conjecture is missing). Further (numerical) examples show that 
violation of $\bf{(C_2)}$ 
with certain parameters imply violation of Eq.(5) (confirming the main claim of I). This fact does not mean that $\bf{(C_2)}$ 
(besides $\bf{(C_1)}$ $\bf{(C_3)}$) is necessary for validity of Eq.(5), $T=0$, since it is possible to derive conditions 
$\bf{(C_1)}$, $\bf{(C^{*}_2)}\ne \bf{(C_2)}$, $\bf{(C_2)}$ that also lead to $T=0$ in Eq.(5). An example is given by
$$
a=\frac{c_1^2}{12c_2},\quad c_3=\frac{16}{9}c_1c_2>0,\quad \bf{(C^{*}_2)}  
$$ 
which defines another subspace $P_{C^{*}}\ne P_C$ of $P$, associated to another family of solutions $\Psi(t, z)$, the so-named 
rational solutions \cite{AkhAT}.

To be specific, condition $\bf{(C_2)}$ implies existence of two simple roots $\{0,\frac{8c_2}{c_1}\}$ and a double root $\frac{4c_2}{c_1}$ 
of Eq.(3) (Fig.11). Condition $\bf{(C^{*}_2)}$ implies existence of a simple root ($h(0)=0$) and a triple root $\frac{4c_2}{c_1}$ (Fig.12). 
In both cases, condition $\bf{(C_1)}$ is satisfied but not necessary (if $\bf{(C_2)}$, simple root $\frac{8c_2}{c_1}$ implies $T\ne 0$, 
as shown in Section IV,   
double root $\frac{4c_2}{c_1}$ implies $f_z(t, z)=0$, treated in I; if $\bf{(C^{*}_2)}$, simple root $\frac{4c_2}{c_1}$ 
leads to $h=constant=\frac{4c_2}{c_1}$ and hence to $f_z(t, z)=0$ again). Any initial condition $h(z_0)$ 
with $0\le z_0<\frac{4c_2}{c_1}$ can be used, leaving $\bf{(C_3)}$ unchanged and leading to the same family of solutions 
as for $h_0=h(0)=0$. Hence, considering Fig.1, an initial condition at  any $z(<0)$ can be used to construct solutions 
$h(z), f(t, z)$ \cite{AkhAT}. -- This fact may be important for modeling physical phenomena by NLSE. 

Since both conditions $\{\bf{(C_1)}, \bf{(C_2)}, \bf{(C_3)} \}$ and $\{\bf{(C_1)}, \bf{(C^{*}_2)}, \bf{(C_3)}\}$ are sufficient for 
$T=0$, with condition $\Delta_h=0$ in common, the somewhat basic problem is whether
$\Delta_h=0$ is necessary for the validity of $T=0$ in Eq.(5). In this respect, considerations seem suitable 
to find further solutions of the NLSE.


\section*{References}



\begin{thebibliography}{99}









\bibitem{AEK}  N. Akhmediev, V.M. Eleonskii and N.E. Kulagin, Theor. Math. Phys., Vol. 72, 809 (1987).














\bibitem{Se} (a) H.W. Sch\"urmann and V.S Serov, Theor. Math. Phys., Vol. 219(1), 557 (2024); (b) H.W. Sch\"urmann and V.S Serov, Theor. Math. Phys., Vol. 219(3), 1060 (2024).











\bibitem{AkhAT} N. Akhmediev, A. Ankiewicz and M. Taki, Phys. Lett. A, Vol. 373, 675 (2009).

\bibitem{Note} A quartic polynomial defines a genus-one algebraic curve. The same curve is described in its canonical form by 
the Weierstrass cubic $Y^2=4y^3-g_2y-g_3$. Thus, invariants and discriminant of the quartic and of the Weierstrass' function 
$\wp(z;g_2,g_3)$ as solution of the cubic are the same (up to positive constant).














 



\end{thebibliography}
\end{document}